\documentclass[pre,aps,twocolumn,amsmath,amssymb,superscriptaddress]{revtex4-1}
\usepackage{color}
\usepackage{amssymb}
\usepackage{esint}

\usepackage{amstext}
\usepackage{graphicx,epstopdf,hyperref}
\usepackage{color}
\usepackage{amsmath}
\usepackage{color}
\usepackage{enumitem}
\usepackage{soul}

\makeatletter

\begin{document}

\title{Transient subdiffusion from an Ising environment}

\author{G. Mu\~noz-Gil}
\author{C. Charalambous}

\author{M.A. Garc\'{\i}a-March}
\email{miguel.garcia-march@icfo.es}
\affiliation{ICFO-Institut de Ci\`encies Fot\`oniques, The Barcelona Institute of Science and Technology, 08860 Castelldefels (Barcelona), Spain}
\author{M.F. Garcia-Parajo}
\affiliation{ICFO-Institut de Ci\`encies Fot\`oniques, The Barcelona Institute of Science and Technology, 08860 Castelldefels (Barcelona), Spain}
\affiliation{ICREA - Pg. Llu\'{\i}s Companys 23, 08010 Barcelona, Spain}
\author{C. Manzo}
\affiliation{ICFO-Institut de Ci\`encies Fot\`oniques, The Barcelona Institute of Science and Technology, 08860 Castelldefels (Barcelona), Spain}
\affiliation{Universitat de Vic - Universitat Central de Catalunya (UVic-UCC), C. de la Laura,13, 08500 Vic, Spain}
\author{M. Lewenstein}
\affiliation{ICFO-Institut de Ci\`encies Fot\`oniques, The Barcelona Institute of Science and Technology, 08860 Castelldefels (Barcelona), Spain}
\affiliation{ICREA - Pg. Llu\'{\i}s Companys 23, 08010 Barcelona, Spain}
\author{A. Celi}
\affiliation{ICFO-Institut de Ci\`encies Fot\`oniques, The Barcelona Institute of Science and Technology, 08860 Castelldefels (Barcelona), Spain}

\begin{abstract}
We introduce a model, in which a particle performs a continuous time random walk (CTRW) coupled to an environment with Ising dynamics. The particle shows locally varying diffusivity determined by the geometrical properties of the underlying Ising environment, that is, the diffusivity depends on the size of the connected area of spins pointing in the same direction. The model shows anomalous diffusion when the Ising environment is at critical temperature.  We show that any finite scale introduced by a temperature different from the critical one, or a finite size of the environment, cause subdiffusion only during a transient time. The characteristic time, at which the system returns to normal diffusion after the subdiffusive plateau depends on the limiting scale and on how close the temperature is to criticality.   
The system also displays apparent ergodicity breaking at intermediate time, while ergodicity breaking at longer time occurs only under the idealized infinite environment at the critical temperature. 
\end{abstract}

\maketitle

\begin{figure*}
	\includegraphics[width=0.95\textwidth]{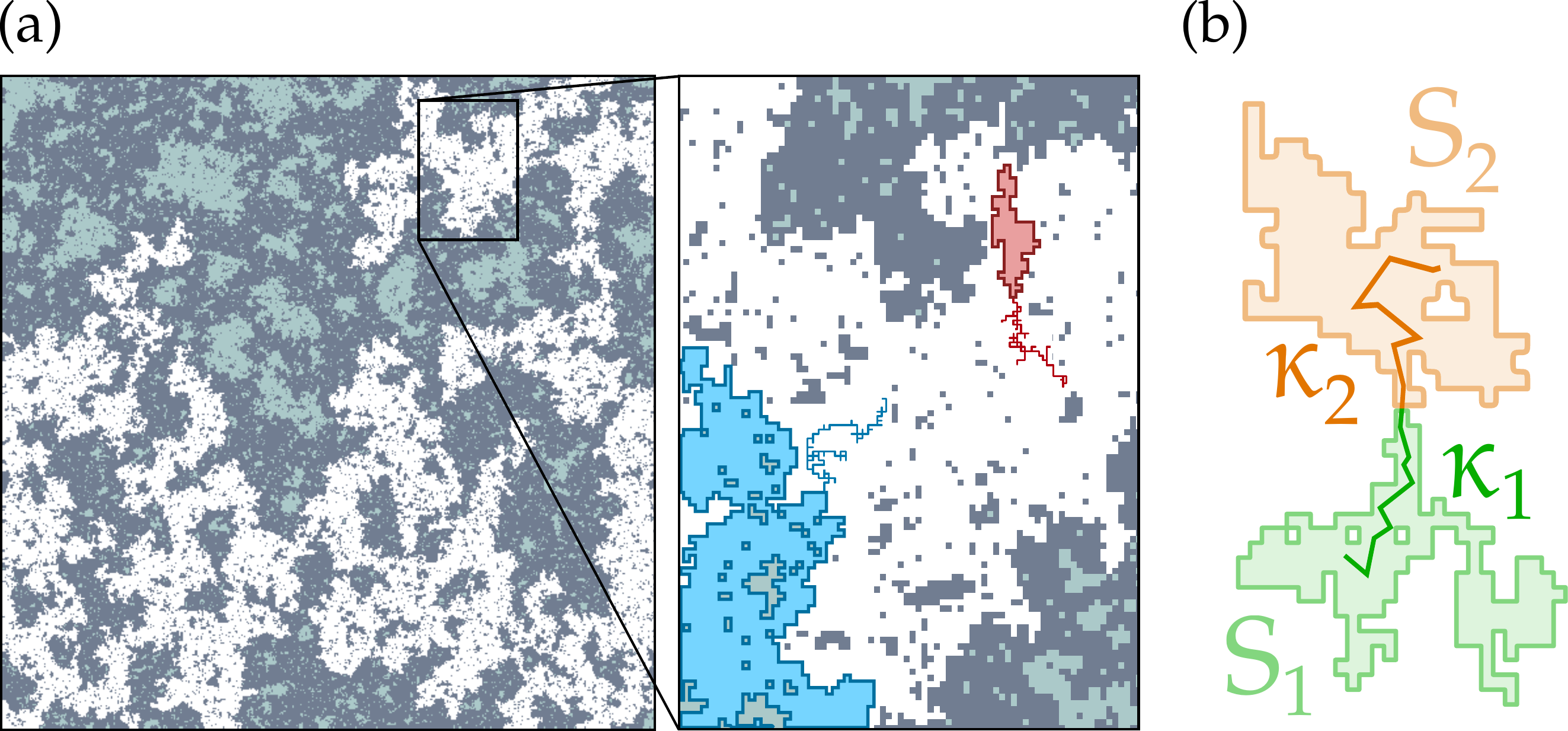}
	\caption{Schematic of the CTRW performed by the particle in an Ising environment. (a) We present a squared Ising lattice close to the critical temperature, $k_B T_c \approx 2.2691 J$, with $k_B$ the Boltzmann constant, and side length of $L=500$. Here dark (light) blue is pointing up (down) spins. The white area is the domain in which the particle sits at time $t$. We show a zoom of a region in which two particles (red and blue curves) perform a CTRW with waiting-time PDF given by Eq.~(\ref{eq:waitingtimekappa}). We highlight the domains entered by the particles after leaving the initial domain with the same color code as the trajectories. (b) An scheme of a particle's motion through two different clusters. Here,  $\kappa$ is determined from the size of the domain in which they sit, according to Eq.~\eqref{eq:transform}. \label{fig0} } 
\end{figure*}

\section{Introduction}
\label{sec:intro}
Deviations from Brownian diffusion - expressed by a nonlinear dependence of the mean square displacement as a function of time - are ubiquitous in Nature, with examples ranging from quantum physics to biology. Although several models have been proposed for its explanation~\cite{2014MetzlerPCCP}, the occurrence of {\it anomalous} diffusion can be generally interpreted as the result of energetic and/or geometric disorder of the environment, in which the motion takes place~\cite{book_klafter}. Biological membranes, because of their fluid nature and the heterogeneous composition of lipids and proteins, have been often used as an ideal test system to demonstrate the occurrence of exotic diffusion and/or the validity of the proposed models. Recent experimental and theoretical advances on non-brownian motion on membranes are reviewed in~\cite{2015Krapf,2016Metzler}. Moreover, in the last years, it is becoming increasingly evident that membranes are not only a key structure of the cell because they confer protection and separation: they also play a major role for signaling by controlling proteins' spatio-temporal organization, which is essential for the regulation of the biological function~\cite{2016Bernardino}.  Therefore, understanding the mechanisms underlying the observed anomalous diffusion in the cell membrane is fundamental to fully characterize cell behavior.

Anomalous diffusion in the plasma membrane can be explained as the consequence of several processes, involving different interactions with molecular components of the surrounding environment~\cite{2015Krapf,2016Metzler}. Examples are transient binding~\cite{2011Weigel}, heterogeneities~\cite{2015ManzoPRX, 2017Weron}, membrane compartmentalization by the actin cytoskeleton~\cite{2017Sadegh} and macromolecular crowding~\cite{2008Dix}. Besides membrane proteins, the lipids - the most abundant components of the cell membrane - contribute to this heterogeneous organization by interacting with cholesterol, proteins and the cytoskeleton in order to partition the membrane~\cite{2010Lingwood}, as a consequence of these dynamic interactions, also lipids exhibit non-Brownian diffusion behavior~\cite{2014HonigmannNatComm}. Due to the presence of miscibility
critical points and associated long-range critical fluctuations, the compositional heterogeneity of the plasma membranes has been often modeled as a nearly critical environment through an Ising model~\cite{2011Machta, 2014HonigmanneLife}. 

A number of models have introduced space- and/or time-dependent diffusion coefficients to explain the anomalous diffusion observed in the plasma membranes, e.g. by diffusion of the particles in patches with random diffusivity~\cite{2014MassignanPRL}. Here, we go a step forward by linking this spatial- or time-dependent diffusion to a structured interacting environment. We introduce an heterogeneous model, in which a particle performs a continuous-time random walk (CTRW) in a critical environment described by an Ising model. That is, at every site of the lattice there is a spin, which can point up or down, and which interacts with its nearest-neighbors following an Ising model.
  The particle dynamics heterogeneity is linked to the geometrical properties of the environment. We let the waiting time distribution depend on the dimension of the domain, i.e. the connected region of spins pointing in the same direction, that the particle visits. The probability of finding a domain of a given size $S$ decays exponentially, when the correlation length of the lattice is finite, while becomes algebraic at criticality. Therefore, the CTRW is in general non-Markovian: when the particle moves, the domain it enters (and thus its diffusivity) depends on the domains it has visited before. The appearance of memory originates from the simultaneous time evolution of the particles and of the Ising environment. Such motions can be disentangled in two extreme and opposite limits, i.e. when the motion of the particle occurs at a timescale much faster or much slower than the evolution of the Ising environment. The latter regime can hold only out of criticality, due to the critical slowing down of the dynamics of the Ising environment, see for instance~\cite{1969HalperinPR, 2005Calabrese}. Here we focus on a particle performing such heterogeneous CTRW in squared two-dimensional lattices, however the model can be easily extended to three or higher dimensions.
  Note that, since the lattice we consider is regular, the situation we describe is not equivalent to that of walks on percolation structures~\cite{book_klafter,book_Stauffer}. The particle dynamics depends on the distribution of domain sizes given by the Ising environment, but not on the geometry of the percolation structure.  

We will show that requiring the existence of a minimal energy scale (as occurs out of criticality) implies that subdiffusion can appear only as a transient behavior, while at longer times the particle experiences normal diffusion. Transient subdiffusion can have different manifestations~\cite{2008Bronstein} and/or appear in very different contexts~\cite{2004Jung,2005Berthier,2007Saxton,2016Spiechowicz,Yamamoto2015,Yamamoto2017}. In~\cite{2008Bronstein}, the total motion of telomeres in the nucleus of mammal cells shows transient subdiffusion, which the authors relate to the reptation model. In~\cite{2004Jung,2005Berthier}, the behavior of diffusion as a function of the temperature is explained in certain glasses by modeling them as dynamical environments giving raise to transient subdiffusion.
 
The paper is organized as follows. Section~\ref{sec:model} describes the model under consideration. In Sec.~\ref{sec:infinite} we discuss the subdiffusive behavior of a particle performing a CTRW in an infinite Ising lattice at critical temperature. In Sec.~\ref{sec:finite} we describe how the departure from such an idealistic model, i.e. finite size of environment or departure away from criticality,  affect the motion of the particle.  In Sec.~\ref{sec:EB} we comment about the ergodicity of such systems, by means of the ergodicity breaking parameter (EB). Finally, we present our conclusions in Sec.~\ref{sec:conclusions}.

\section{Description of the Model}
\label{sec:model}

We consider a particle moving in a hypercubic lattice of dimension $d$ and side length $L$. For simplicity, we describe in details the two dimensional case, $d=2$, but the result can be easily extended to three or higher dimensions. The particle performs a CTRW in this lattice, with equal probability of moving to each of the neighboring sites. The unitary displacements are done at stochastic times given by the waiting-time probability distribution function (PDF)
\begin{equation}
\label{eq:waitingtimekappa}
\phi_\kappa(t)=\frac{1}{\kappa\lambda}\exp\left[-t/\kappa\lambda\right],
\end{equation}
where $\lambda$ represents a time scale, while the number $\kappa$ accounts for changes in diffusivity induced by properties of the environment.

Generally speaking, we consider models, where spatio-temporal properties of the environment produce different values of $\kappa$, thus giving rise to a probability distribution $P(\kappa)$. From a mathematical point of view, in order to determine the motion performed by the particle,  it is necessary to calculate the convolution of the waiting time distribution in Eq.~(\ref{eq:waitingtimekappa}) with the probability distribution $P(\kappa)$. The functional form of $P(\kappa)$ thus plays a major role in determining the kind of motion experienced by the particle.  In particular,  $P(\kappa)$'s corresponding to a $\delta$ distribution or decaying exponentially on a small scale produce normal diffusion. In contrast, a power law PDF, $P(\kappa)\sim\kappa^{-\sigma}$ with $1<\sigma<2$, displaying infinite mean and variance, produces anomalous subdiffusion equivalent to the quenched trap model.

In the simplest example, we consider an environment obtained as a system of classical spins, $s_{ij}=\pm1$, which occupy the same square lattice on which the particle performs the CTRW and  
interact with the two dimensional Ising Hamiltonian
\begin{equation}
H_{\mathrm{Ising}}=-J\sum_{\langle i j \rangle} s_i\,s_j.
\end{equation}
Here $\langle i j \rangle$ are the nearest-neighbors pairs in the lattice.   
Close to criticality, the Ising system displays an arrangement in domains (defined as regions of nearby spins pointing in the same direction) with a power law distribution of sizes. As commented in Sec.~\ref{sec:intro}, such distribution has been used to describe the heterogeneities  present in plasma membranes \cite{2011Machta,2014HonigmanneLife}. In this context, relating $P(\kappa)$ to such distribution of sizes provides a way to connect spatio-temporal properties of the environment at criticality to the stemming of subdiffusion, and thus offers an analytical tool to study their interplay.

To this aim, we consider a putative microscopic model for which the number $\kappa$ is a function of the size $S$ of the domain that the particle visits at each step through  
\begin{equation}\label{eq:transform}\kappa=S^{\eta}. \end{equation} 
As shown in Fig.~\ref{fig0} the particle performs a CTRW in an Ising environment. For a number of steps the particle stays in the same domain (white area of panel (a), green area of panel (b)), and since $\kappa$ is determined by the size of the domain through Eq.~\eqref{eq:transform}, it remains  constant for all these steps. When the particle leaves this domain, it enters a new one of size $S'$ (blue and red areas in panel (a), orange area in panel (b)), hence taking a new value of $\kappa$ which is kept until the particle exits that domain. In Fig.~\ref{fig0} we consider that the Ising environment is at equilibrium and does not evolve at all. In the next section we discuss how we introduce dynamics for the Ising environment. 
We note here that we assume that only the size of the domain matters in the determination of $\kappa$, irrespectively of the spin direction in that domain.  

A key element to characterize the motion of the particle in these disordered environments is to find the probability $P(S,j)$ that at a step $j$ the particle sits in a domain of size $S$. This probability depends both on the frequency of a domain of size $S$ in the Ising system, $P(S)$, and on the domain occupied by the particle in its previous steps. 
The frequency of a domain of size $S$, $P(S)$, follows the behavior 
\begin{equation}
\label{eq:distrS}
P(S)\propto S^{-\tau}\exp[-S/S^*],
\end{equation} 
with the critical parameter $\tau=187/91\approx2.05$~\cite{1967Fisher,book_Stauffer} and with $S^*$ representing the characteristic size of a domain. This distribution is valid for large enough $S$. Through Monte Carlo simulations of two dimensional Ising lattices near critical temperature and lattice length $L=200, 500$ and $1000$ spins, we numerically checked  that Eq.~(\ref{eq:distrS}) describes well the distribution of sizes even for small $S$ (see Appendix~\ref{sec:appA}). Thus, for the sake of simplicity, we consider it valid for $S\geq1$. 

Based on the previous discussion, it is possible to make some general considerations regarding the model.  First, it is worth noticing that, since the minimum domain size is imposed by the lattice to be $S=1$, one cannot use the conventional approximation of a probability $P(S)$ decaying to zero for small $S$, as it was done, e.g., in Refs.~\cite{2016Charalambous,2001BeckPRE}. Second, since the characteristic size of a domain $S^*$, determined by the correlation length~\cite{1967Fisher,book_Stauffer}, diverges at criticality, $P(S)$ displays an algebraic behavior at the critical temperature.  Away from criticality and/or for finite-size environments, the correlation length, instead, has a finite value, thus establishing an upper cutoff to $P(S)$, corresponding to the maximum domain size as seen from Eq.~(\ref{eq:distrS}) and Ref.~\cite{1996Cardy}. 

Third, $P(S,j)$ depends strongly on the previous history of the system $P(S,j|S',j-1;S'',j-2;\cdots)$. This non-Markovian behavior is a consequence of the simultaneous temporal evolution of the particle position and of the environment. These contributions can be disentangled in two opposite limits, corresponding to the cases, in which the particle motion is much faster or much slower than the dynamics of the environment. We discuss this problem in detail in next section. 

Again, we would like to point out that, although we focus our discussion to two dimensional lattices, our results can be extended to larger dimensions by substituting the corresponding value of $\tau$ in~(\ref{eq:distrS}). For instance, in three dimensions we have $\tau=2.53$, see Ref.~\cite{book_Stauffer}. In this case the CTRW of the particles will occur in cubic lattices, and the domain sizes will be volumes instead of areas. The rest of the discussion can be applied directly.

 \section{Infinite Ising environment at criticality}
 \label{sec:infinite}

We first consider our model in the limit of an infinite environment close to the critical temperature,  
where the simultaneous evolution of the environment and the particle set different possible scenarios. If the particle moves much faster than the environment, or in other words, the time scale $\lambda$ in Eq.~\eqref{eq:waitingtimekappa}  is sufficiently shorter than the dynamical critical time $\varsigma$ of the environment (see Sec. \ref{sec:finite}),  one can consider that the environment is not evolving at all.  
Then, if the particle reaches a domain of size $S$, it will perform a number of steps before entering into another domain. This is encoded in the second term in the probability $P(S,j)$. Due to the non-Markovian nature of the system under consideration, it is difficult to obtain an analytical expression for this term. To treat this issue in the simplest way, we assume that once a particle reaches a domain of size $S$, it performs 
$n$ steps 
in that domain, 
starting and ending on the domain's boundary (see Fig.~\ref{fig0}). In the current limit (fast particle, the domains do not evolve at all), the number of steps $n$ depends on the size $S$ of the domain, but also on the domain shapes and on the initial position of the particle.  The average number of steps $\overline n$ to reach the boundary must necessarily be a monotonically growing function of the domain size $S$. In the simplest approximation of a domain of  circular shape, the particle will perform  on average $\overline{n} \propto S^{1/2}$ steps before leaving a domain of size $S$.  By means of numerical calculations, we find that the average number of steps $\overline n$  in the domains of a two dimensional Ising lattice grows as $\sim S^{0.3}$ (see Appendix~\ref{sec:appA}).
 
In the scenario of the domains having dynamics, the average number of steps performed by a particle in the same domain will change, and in general will   be smaller for a fast motion.
We can relax the relation between the average number of steps and the domain size by taking $\overline{n} \propto S^{\mu}$, with $0\leq\mu\leq0.3$.  
Here, 
$\mu$ can be considered as a free parameter that accounts for the effect of a dynamical environment.  Therefore, the tuning of $\mu$ thus allows us to treat the different time scale dynamics of the environment and the particle. 

From previous statements, we can now give a full description of the probability of the particle entering a domain with size $S$. This probability has to take into account the distribution of the Ising model $P(S)$, the probability of finding that domain in an square environment of length $L$ and the average number of steps made on it
\begin{equation}
\label{eq:distrSn}
P_{\rm{new}}(S) \propto P(S)\cdot \frac{S}{L^2}\cdot S^\mu\propto S^{-\zeta+1+\mu},
\end{equation}
where the latter equality, obtained by means of Eq.~(\ref{eq:distrS}), holds at criticality with $\zeta=\tau-1=1.05$.

 We can now use the transformation in Eq.~(\ref{eq:transform}) to express the probability of sampling a value of $\kappa$  from the new domain as 
 \begin{equation}
 \label{eq:distrK} 
 P_{\rm {new}}(\kappa)=  P_{\rm {new}}(S) \frac {dS}{d\kappa} \propto \kappa^{-\sigma(\mu)}, \\
 \end{equation}
 with 
 \begin{equation}
 \sigma(\mu)=\frac {\zeta-1-\mu}{\eta}+1.
 \label{eq:beta}
 \end{equation}
 
For $\mu<\zeta-1$, the dynamics of the environment is fast enough for the particle to be able to sample completely the distribution $P(S)=S^{-\xi}$. In contrast, for $\mu\ge\zeta-1$, the probability of finding bigger $\kappa$ increases exponentially. Therefore there is a high probability of entering in a sufficiently big domain and remain there for a large time $t\rightarrow\infty$, hence not visiting the rest of the domains. As a consequence, it is necessary to introduce a cut-off to the distribution $P(\kappa)$, not only to prevent this ``confinement'', but also to transform it into a normalizable probability distribution function. In short,  distribution $P_{\rm{new}}(\kappa)$ is well defined for $\mu <   \zeta-1$ and we  normalize it considering a cut-off at large $\kappa$ for $\mu >   \zeta-1$. 
In section~\ref{sec:finite}, we will discuss in detail the effect that such a cut-off has on the diffusion. 


  With this,  we can convolute $P_{\rm{new}}(\kappa)$  with Eq.~(\ref{eq:waitingtimekappa}) in order to determine the waiting time distribution of the particle. We will focus on the 
the case with $\mu<   \zeta-1$, and leave considerations on the case with a cut-off for section~\ref{sec:finite}. The convolution is obtained as follows
\begin{align}
\label{eq:psiinf}
\psi(t)=&\int^{\infty}_{1} P_{\rm{new}}(\kappa)\phi_\kappa (t)d\kappa\\ 
&= \nonumber \frac{1}{\lambda}\left(\frac{t}{\lambda}\right)^{-\sigma(\mu)} \left(\Gamma[\sigma(\mu)] - \Gamma[\sigma(\mu), t/\lambda]\right),
\end{align}
where $\lambda$ represents a characteristic time scale and $\Gamma[\cdot]$ ($\Gamma[\cdot,\cdot]$) is the complete (upper incomplete) Gamma function.
From here one can calculate the  ensemble-averaged mean squared displacement (eMSD)  $M_2(t)=\langle x^{2}\left(t\right)\rangle$, where $x$ is the distance of the particle from its initial position and $\langle\cdot\rangle$ represents the average over multiple trajectories. The eMSD is~\cite{2016Charalambous} 
\begin{equation}
 M_2(t) \approx \frac{1}{\Gamma(\sigma(\mu))}\!\left(\frac{t}{\lambda} \right )^{\alpha(\mu)}\!\!,
 \label{eq:eMSD}
\end{equation}
with $\alpha(\mu)=\sigma(\mu)-1$. The last relation gives us insight on which kind of particle-environment dynamics will show anomalous diffusion. The exponent of the eMSD, $t\partial_t M_2(t)$,  fulfills the condition for subdiffusive motion $0<\alpha(\mu)<1$ only for $1<\sigma(\mu)<2$ . Therefore,  there is a limited number of combinations of the dynamic parameters $\mu$ and $\eta$ that produce a subdiffusive behavior [see Eq.~(\ref{eq:beta})].

We perform a series of Monte Carlo simulations to support our analytical results. These consist in introducing a particle in a two dimensional Ising lattice, which - through a random walk - explores the distribution of domains~(\ref{eq:distrSn}), thus performing the motion whose mean square displacement is predicted by Eq.~(\ref{eq:eMSD}). However, as with many other complex systems, an exact simulation of the previous behavior has a high computational cost and introduces a finite-size effect to the Ising lattice. In order to reproduce the movement of the particle in such a system, we propose the following prescription: (i) a value of $S$ is randomly selected from the distribution $P_{\rm{e}}(S)=S^{1-\zeta}$, which takes into account only the contribution of the environment, in contrast to Eq.~(\ref{eq:distrSn}); (ii) the corresponding $\kappa$ is calculated with Eq.~(\ref{eq:transform}); (iii) for each of the $\bar{n}=S^{\mu}$ steps performed in the domain, a different waiting time is retrieved stochastically from the distribution~(\ref{eq:waitingtimekappa}). The implementation of Eq.~(\ref{eq:waitingtimekappa}) and $P_{\rm{e}}(S)$ into a computational code is outlined in appendix~\ref{sec:numimple}. Using this prescription, we numerically obtained the eMSD. 

\begin{figure}
\includegraphics[width=0.95\columnwidth]{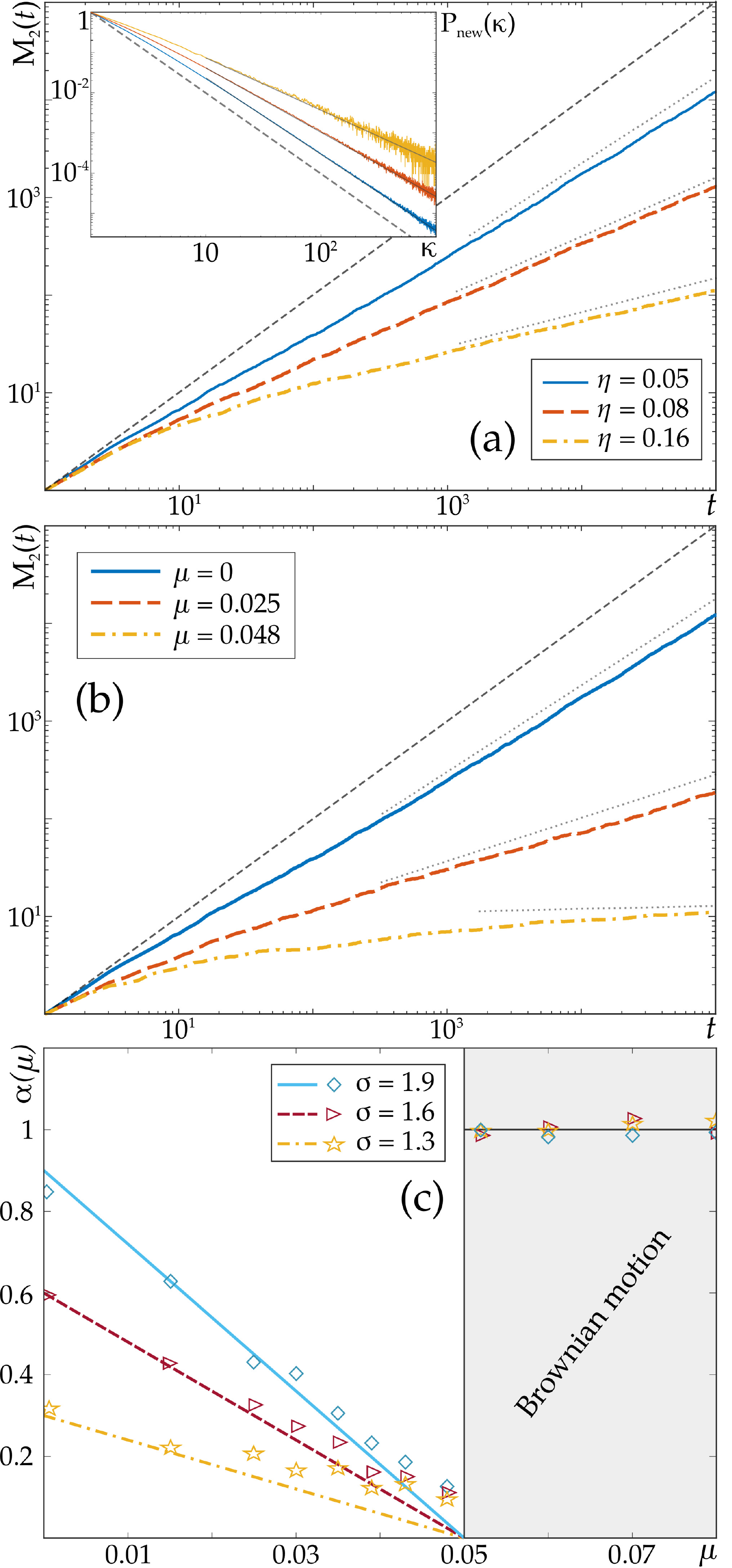}
\caption{Ensemble-averaged mean squared displacement (eMSD): comparison between the theoretical predictions and Monte Carlo simulations. (a) eMSD for different theoretical values of $\eta$ and $\mu=0$ for infinite size Ising environments at critical temperature. The movement is subdiffusive, that is ${\rm M}_2 (t) \propto t^{\alpha}$, with $0\leq\alpha\leq1$. Dashed black line represents the Brownian motion limit. Dotted lines show the theoretical prediction of Eqs.~(\ref{eq:beta}) and~(\ref{eq:eMSD}) (same for (b)). Inset: The corresponding probability distribution function $P_{\rm{new}}(\kappa)$ used for the simulations. (b) eMSD obtained for different values of $\mu$ and $\eta=0.05$. The expected subdiffusive behavior is observed. 
(c) Relation between the eMSD exponent $\alpha(\mu)$ and the time scale parameter $\mu$. In this panel, symbols represent the numerical calculations while  lines are theoretical predictions. We see that for $\mu>\xi-1=0.05$ normally diffusing Brownian motion occurs, i.e. $\alpha(\mu)=1$.  \label{fig1} }
\end{figure}

 In Fig.~\ref{fig1} we show numerical calculations for the eMSD for different values of $\sigma(\mu)$. As commented before, the two parameters $\eta$ and $\mu$, corresponding to different particle/environment behaviors, can be adjusted in order to get subdiffusive dynamics. In the case of Fig.~\ref{fig1}(a), we keep $\mu$ fixed and report simulations corresponding to different values of $\eta$ as shown in Eq.~(\ref{eq:transform}). For Fig.~\ref{fig1}(b) we consider the opposite case, in which  $\eta$ is fixed and we plot results for different dynamics of the environment. Figure~\ref{fig1}(a) and the corresponding inset show that the exponent obtained for the eMSD in the long time limit is obtained from the exponent of $P_{\rm{new}}(\kappa)$ as  $\alpha(\mu) = \sigma(\mu)-1$. 

As we mentioned previously, there is a critical value of  $\mu=\zeta-1$ at which the particle motion $P_{\rm{new}}(\kappa)$ is no longer normalizable. 
This behavior is illustrated in Fig.~\ref{fig1}(c), in which the exponent of the eMSD $\alpha(\mu)$ decreases linearly with $\mu$ starting in an initial value given by the  parameter $\eta$. Close to the critical value of $\mu=\zeta-1$, the exponent $\alpha(\mu)$ departs from the theoretical prediction (given by Eq.(\ref{eq:beta}) and~(\ref{eq:eMSD})). This is because for $\alpha$ close to zero, the particle does not move for very long periods of time, resulting on a bad sampling of the probability distribution $P_{\rm{new}}(\kappa)$.  Numerically, we find that for  $\mu>\zeta-1$ the particle becomes Brownian, i.e., normally diffusing. In this case, it is necessary to introduce a cut-off to normalize $P_{\rm{new}}(\kappa)$. We discuss in next section that this leads  to normal diffusion in the long time limit.   

\section{Finite size Ising environment and deviations from the critical temperature}
\label{sec:finite}

In the previous section, we considered a random walk in an environment corresponding to an infinite Ising lattice close to criticality. However, it must be observed that an infinite lattice at critical temperature experiences a critical slowing down. The correlation between a spin value at a lattice site at time $t$ and its value at the initial time $t=0$ is given by
\begin{equation}
\label{eq:criticalslowingdown}
 \langle s_i(t)\,s_i(0) \rangle\propto \exp [-t/\varsigma],
\end{equation}
where $\varsigma$ is the dynamical critical time, which diverges at the critical temperature. As stated in Sec.~\ref{sec:infinite}, the comparison of the critical time to $\lambda$ from Eq.~\eqref{eq:waitingtimekappa} gives the relation between the dynamics of the environment and the particle. The limit where $\varsigma \ll \lambda$ implies that the environment evolves much faster than the particle, making the pattern the latter is seeing at one step completely uncorrelated to the one of the next step. When $\varsigma \gg \lambda$, we recover the opposite limit, in which the particle moves through an environment with very small dynamics. The dynamical critical scaling hypothesis~\cite{1976Ma,1969HalperinPR,1977HohenbergRMP,1977SuzukiPTP} states that the dynamical critical time $\varsigma$ behaves as 
\begin{equation}
\label{eq:criticalslowingdown1}
 \varsigma\propto \xi^z,
\end{equation}
with $z$ being the dynamical critical exponent, calculated numerically to be $z=2.167$~\cite{1996NightingalePRL,2016LinPRE}. The correlation length $\xi$ diverges as $\xi=\left|T-T_{\rm crit}\right|^{-\nu}$, with $\nu=1$. Therefore, the critical relaxation time $\varsigma$ diverges at critical temperature, giving infinite time correlations between spins - see Eq.~(\ref{eq:criticalslowingdown}). Equations~(\ref{eq:criticalslowingdown}) and~(\ref{eq:criticalslowingdown1}) thus imply that, at critical temperature and infinite size environment, one cannot find a time scale quick enough as to assume that -- each time the particle moves -- a different pattern of domains is sampled.

However, in practice the temperature will depart from criticality or the lattice will have a finite size $N=L^2$. For finite systems at critical temperature, the relaxation time is finite, $\varsigma\propto L^z$. Similarly, in the infinite size case away from criticality $\varsigma$ is finite  because the correlation length $\xi$ is finite [see Eq.~(\ref{eq:criticalslowingdown1})]. The time evolution of the particle is thus build up from many exponential waiting times of the form given by Eq.~(\ref{eq:waitingtimekappa}). These have characteristic times $\langle\kappa \lambda\rangle$, the smaller of which is  $\langle\lambda\rangle$. Therefore, the assumption of very slow dynamics of the particle when compared to the Ising environment  translates into $\lambda\gg L^z $ in the case of finite size environment, and $\lambda\gg \xi^z $ in the case of non-critical environment. 
        
The effect of the exponential decay [see Eq.~(\ref{eq:distrS})] in the case of finite environment or deviations from the critical temperature is to introduce a maximum value for the domain size, which is now dictated by the size of the environment $N$ and $S^*$, respectively, since in the case of finite environment, one cannot find larger domains than $N$, while $S\gg S^*$ are exponentially rare in the off-critically case~\cite{book_Stauffer}. Only domains with $S<N$ or $S<S^*$ contribute to the distribution, and for these domains the distribution is effectively $\propto S^{-\tau} $. 

To get a  better understanding of the previous considerations let us consider that the probability distribution Eq.~(\ref{eq:distrK}) is modified as follows
\begin{equation}
\label{Eq:DistrFS}
P(\kappa) = 
   \begin{cases} 
      \propto \kappa^{-\sigma(\mu)}            &\mathrm{if }  \   \ \kappa<   \overline \kappa\\
      0  &\mathrm{if }  \  \  \kappa>   \overline \kappa,
   \end{cases}
\end{equation}
where the cutoff $   \overline \kappa$ is given by $   \overline \kappa=\    \overline S^\eta$ (see appendix~\ref{sec:numimple}).
Here, Eq.~(\ref{Eq:DistrFS}) can be interpreted as the distribution obtained for a finite size environment or a simplified way of modeling departure from the critical temperature given by Eq.~(\ref{eq:distrS}). In the following, we consider that the cutoff in $\kappa$ is due to a finite size in the lattice, and thus to the maximum possible value of $\bar{S}=N$. The insertion of a cutoff is also valid to solve the motion for the case $\mu\ge\zeta-1$, where the probability distribution $P_{\rm{new}}(\kappa)$ cannot be normalized (see Sec.~\ref{sec:infinite}).
Following Eq.~(\ref{Eq:DistrFS}), the integral in Eq.~(\ref{eq:psiinf}) can be written as 
\begin{eqnarray}
\label{eq:psiK}
\psi(t)=&\int^{ \overline \kappa}_{1} P_\kappa(\kappa)\psi_\kappa (t) d\kappa\\
=&\frac{1}{\lambda}\left ( \frac{t}{\lambda}\right )^{-\sigma}\left [\Gamma(\sigma, \frac{t}{\overline\kappa \lambda})-\Gamma(\sigma,\frac{t}{\lambda}) \right ]\nonumber.
\end{eqnarray}  
We can notice that -- since $\Gamma[\sigma, 0]=\Gamma[\sigma]$ -- Eq.~(\ref{eq:psiK}) converges to Eq.~(\ref{eq:psiinf}) for $\overline\kappa\to\infty$. On the basis of the timescales involved, we can identify two temporal regimes: (I) $\lambda \ll t \ll  \overline\kappa \lambda$, and (II)   $ t>  \overline\kappa \lambda$. In the first regime, the times are very large but still smaller than the cutoff in $\kappa$.  In this limit,  $\Gamma(\sigma,\frac{t}{\lambda})$  tends exponentially to zero and we can neglect its contribution. Contrarily, we can consider that $\Gamma[\sigma, \frac{t}{\overline\kappa \lambda}]=\Gamma[\sigma, 0]=\Gamma[\sigma]$ . Therefore, the eMSD will behave as in Eq.~(\ref{eq:eMSD}). In regime (II), that is, when time is larger than all the timescales $\lambda$ and $\overline\kappa$, we expect that both Gamma functions tend exponentially to zero, thus giving normal diffusion at long times.  

\begin{figure}
\includegraphics[width=0.95\columnwidth]{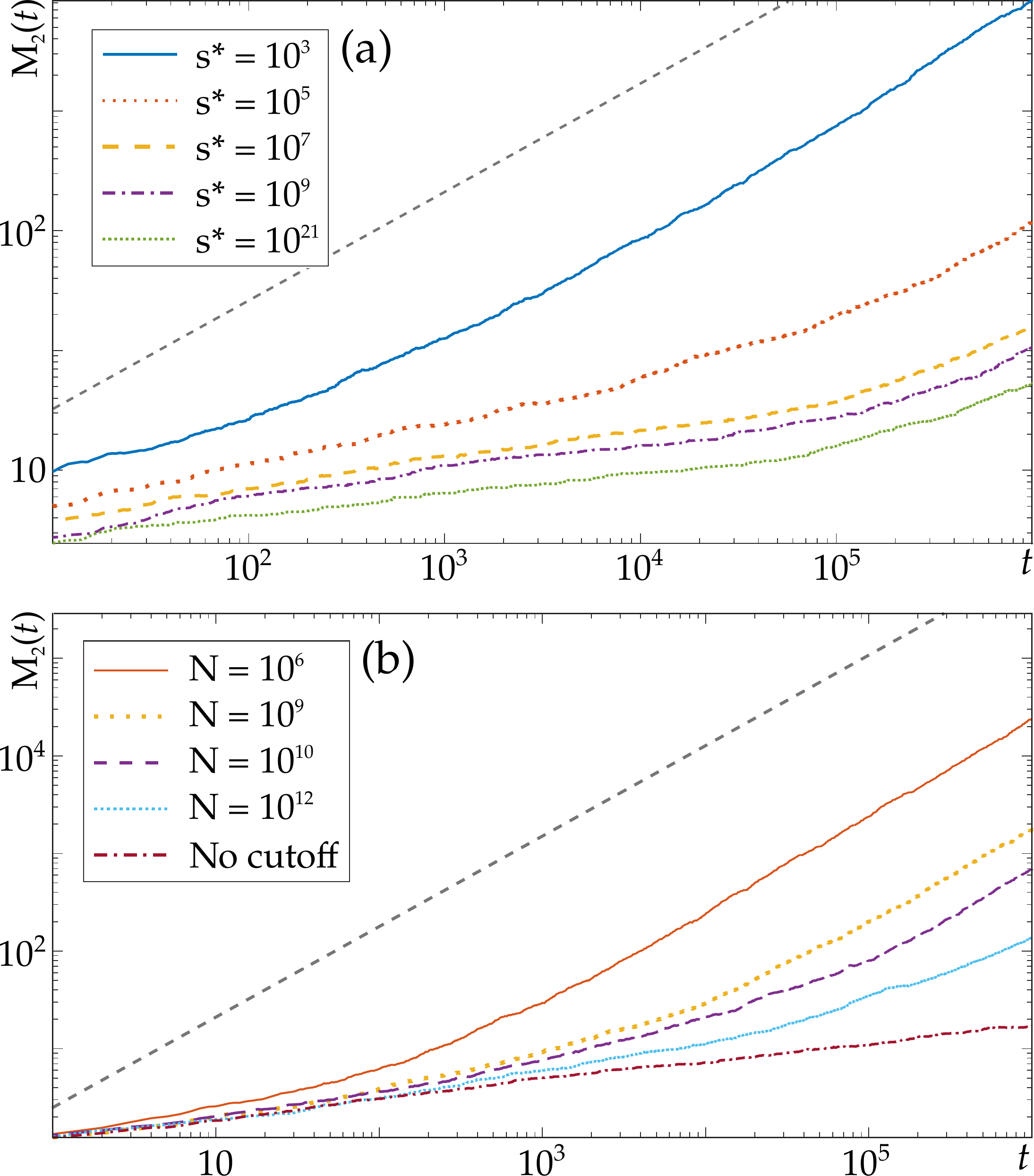}
\caption{Ensemble-averaged mean squared displacement out of criticality due to deviations from the critical temperature and finite size effects. (a) eMSD for $\alpha=0.05$ and different deviations from the critical temperature, described by the typical size $S^*$.  A transient subdiffusive behavior occurs for intermediate times. As $S^*$ is increased, the onset of  diffusive dynamics occurs at a longer time. (b) eMSD for 
$\alpha=0.2$ and different values of the size $N$ of the Ising environment at critical temperature. As shown the movement  is diffusive in the long term behavior. In intermediate times, the finite size cases show the same subdiffusive behavior than the infinite size case (bottom curve). The position at which it departs from subdiffusion is larger as the size is increased. Dashed lines show the Brownian motion limit.  \label{fig4} }
\end{figure}

In Fig.~\ref{fig4}(a) we show the numerically calculated eMSD for different values of $S^*$, when  $\alpha=0.05$. For finite  values of $S^*$, a subdiffusive  plateau occurs at intermediate times. As  $S^*$ is increased, thus getting closer to the critical temperature, one gets a larger subdiffusive plateau.
In Fig.~\ref{fig4}(b) we show the eMSD for the infinite and different finite size Ising environments. The time, at which the eMSD departs from the slope corresponding to the infinite size environment and tends to slope equal to one, is shorter as the lattice size decreases.

\section{Ergodicity breaking}
\label{sec:EB}

The occurrence of anomalous diffusion as a consequence of a motion in a disordered environment is often associated to weak ergodicity breaking~\cite{2014MetzlerPCCP}. 
The presence of nonergodicity implies that the time-averaged mean squared displacement (tMSD) 
\begin{equation*}
\overline{\delta^2}(t,t_{\mathrm{lag}})=\frac{\int_0^{t-t_{\mathrm{lag}}} \left[x(t'+t_{\mathrm{lag}})-x(t')\right]^2 dt'}{t-t_{\mathrm{lag}}},
\end{equation*}
calculated over multiple trajectories of individual particles remains a random variable, and hence irreproducible even in the long time limit.
A quantitative measure of the nonergodicity is thus given by the so-called ergodicity breaking parameter EB
\begin{equation}
\label{eq:EB}
\overline{\mathrm{EB}}=\lim_{t\to\infty}\mathrm{EB}(t)=\lim_{t\to\infty}\frac{\langle(\overline{\delta^2})^2\rangle-\langle\overline{\delta^2}\rangle^2}{\langle\overline{\delta^2}\rangle^2},
\end{equation}
which tends to zero for an ergodic process and is independent of $t_{lag}$~\cite{2008HePRL}. For a CTRW with heavy-tailed waiting time distribution (as in Eq.~(\ref{eq:psiinf})),  the $\overline{\mathrm{EB}}$ parameter  is~\cite{2008HePRL} 
\begin{equation}
 \overline{\mathrm{EB}}=\frac{2\Gamma^2[\sigma]}{\Gamma[2\sigma-1]}-1.
 \label{eq:EB_value}
\end{equation}
Therefore, we used our simulations to determine the tMSD for several particles and thus verify the occurrence of ergodicity breaking in our model. 

\begin{figure}
\includegraphics[width=0.95\columnwidth]{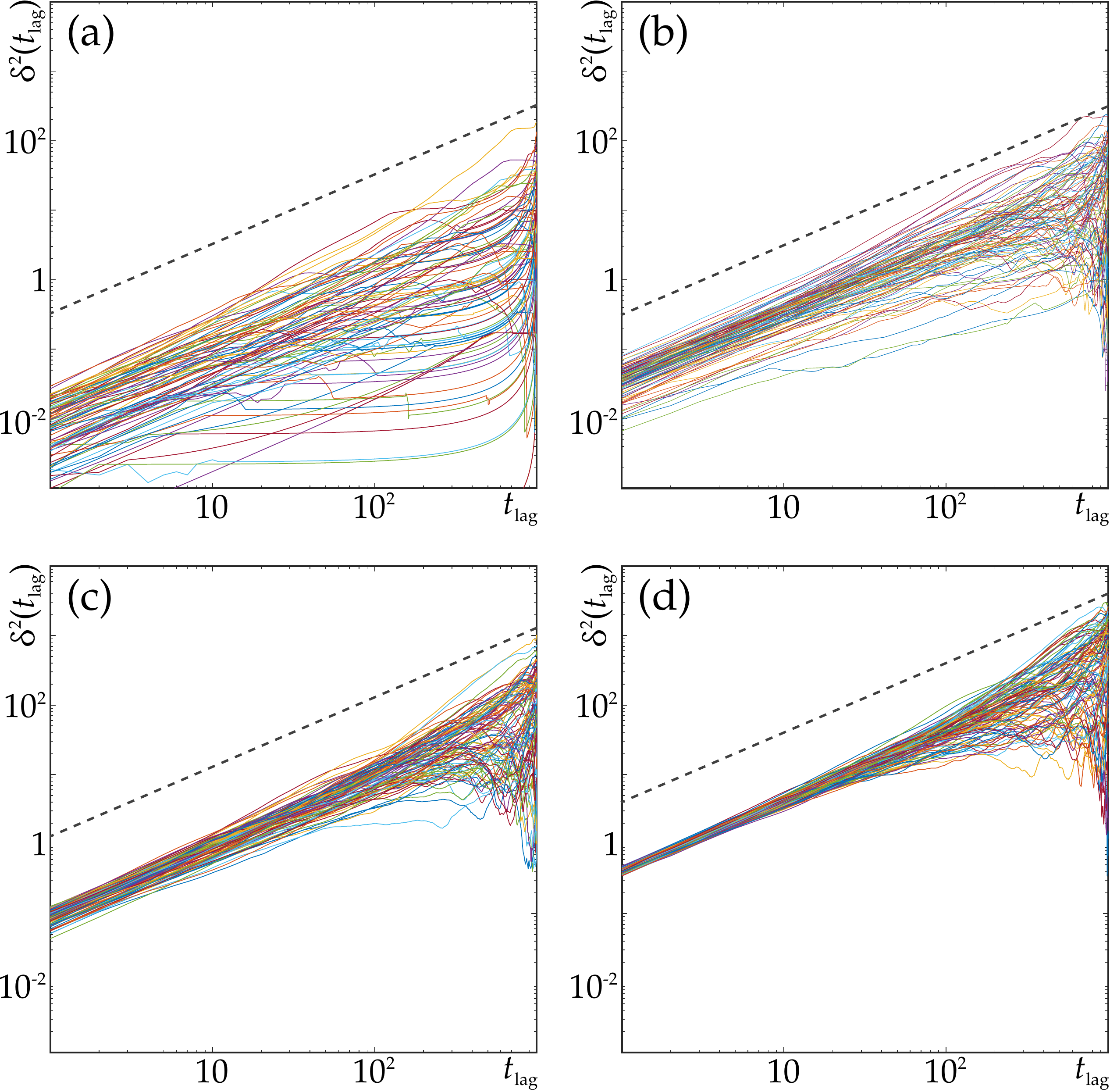}
\caption{ Time-averaged mean squared displacement $\overline{\delta^2}(t,t_{\mathrm{lag}})$ in Ising environment of different sizes. (a) Time-averaged mean squared displacement for $\alpha=0.4$ in the infinite size Ising environment.  As shown, the tMSD remains a random variable in time, as seen by the large scattering observed at all time lags.  (b)-(d) Same in the presence of a maximum size $N=L^2=10^{16}, 10^{12}$ and $10^{6}$, respectively.  \label{fig2} }
\end{figure}

\begin{figure}
\includegraphics[width=0.95\columnwidth]{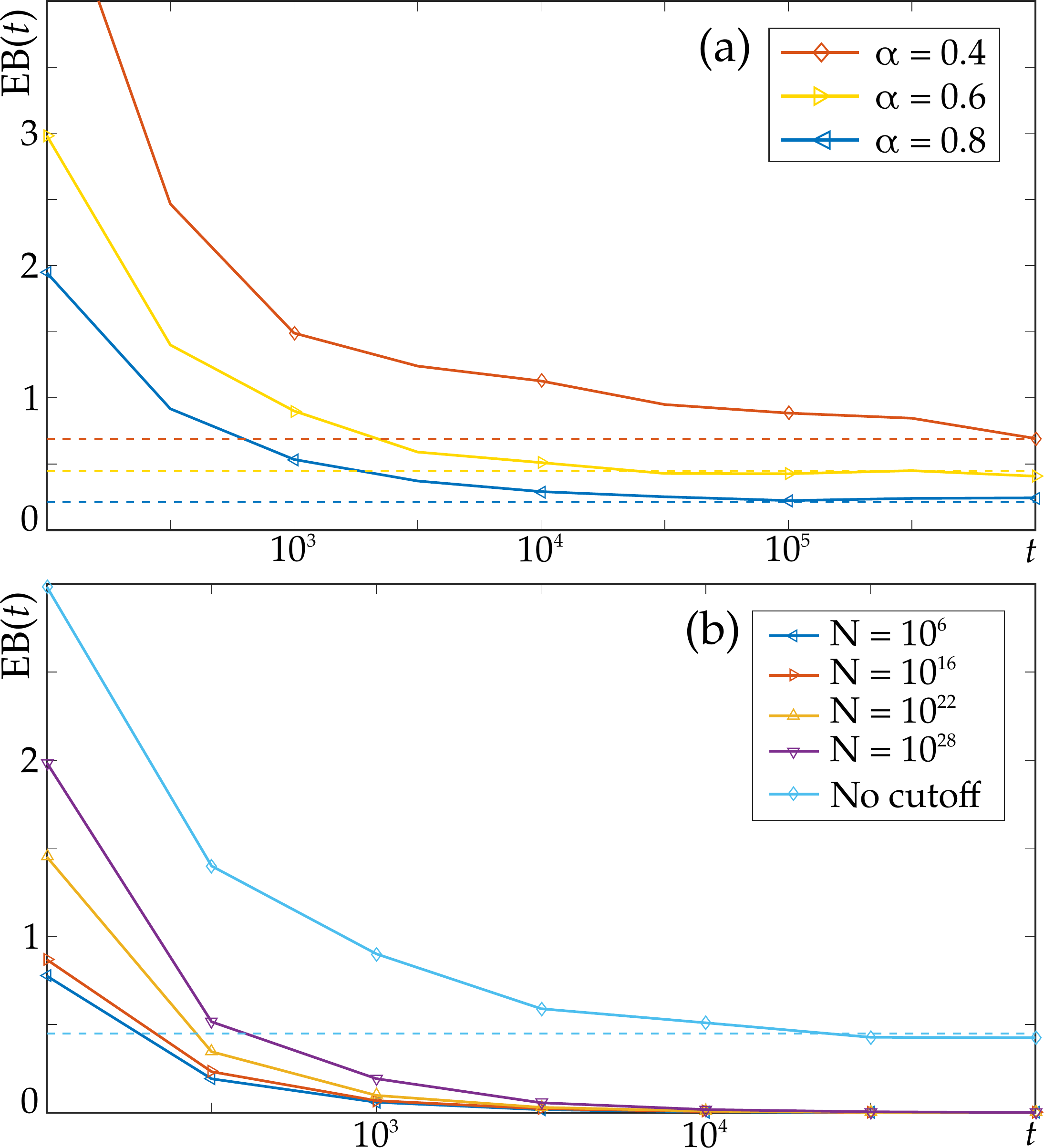}
\caption{Behavior of the ergodicity breaking parameter $\mbox{EB}(t)$ for different processes and sizes of the environment. (a)  Ergodicity breaking parameter for different values of $\alpha$ in the infinite size environment case. This parameter tends asymptotically to the value predicted by Eq.~(\ref{eq:EB_value}) [we indicate the asymptotic value of EB for each $\alpha$ with horizontal lines]. (b)  Ergodicity breaking parameter for $\alpha=0.4$ both for the infinite and finite size environment  cases. For  all finite cases the curves tend to zero asymptotically for the maximum measurement time taken here, $t=10^6$. This means that the behavior is ergodic at long times for finite size if the measurement time is large enough. The time at which it reaches the zero value is larger as the size of the environment is  increased.   \label{fig3} }
\end{figure}

In the fast environment limit at criticality and infinite size system, as shown in Fig.~\ref{fig2}(a), we found that the tMSD exhibits linear behavior at short time lags. Moreover, tMSD curves corresponding to different trajectories are largely scattered, as expected for nonergodic dynamics. This observation is quantitatively reflected in the  value of the EB$(t)$ parameter measured at long times [see Fig.~\ref{fig3}(a)], which moreover tends to the corresponding nonzero values given by Eq.~(\ref{eq:EB_value}) (dashed lines), thus in full agreement with the theoretical prediction. 
The EB$(t)$ parameter gives also a good tool to study deviations from criticality studied in the previous sections. We see in fact that as the particle departs away from the subdiffusive behavior due to finite size and off-criticality effects, the ergodicity of the system increases. 
We plot in  Fig.~\ref{fig2}(b)-(d) the tMSD for $\alpha=0.4$  and different maximum size, and we compare it with the infinize size Ising environment Fig.~\ref{fig2}(a). 
The plots show the dispersion is decreased as the size of the lattice is reduced. Also, the $\overline{\mathrm{EB}}$ reached asymptotically for all finite size cases is zero, while the time at which this value is reached is longer as the environment is made larger [see Fig.~\ref{fig3}(b)]. Thus, for all finite size environments, no ergodicity breaking is predicted asymptotically. 

\section{Conclusions}
\label{sec:conclusions}

We provide a model to describe the effect of critical fluctuations such as those occurring in plasma membranes.  We  link these fluctuations to diffusion heterogeneity observed for the motion of several membrane receptors. In our model, the complex environment with critical fluctuations is exemplified by a lattice of spins evolving with the Ising Hamiltonian close to criticality. Particles diffuse in such environment with locally varying diffusivity, determined by the size $S$ of the visited domains of aligned spins. We show that a particle coupled to an environment with critical behavior displays subdiffusive dynamics. However, in realistic systems, this subdiffusion occurs only transiently due to i) a finite scale introduced by the environment through its finite size or deviations from critical temperature; ii) an intrinsic scale that must be introduced in order to obtain a normalizable distribution of  $\kappa$'s from the critical property, through which the coupling to the environment occurs.
The characteristic time, at which the system goes back to normal diffusion depends on these limiting scales.  Together with anomalous diffusion, the system displays also apparent ergodicity breaking at intermediate time.

We also note that the introduction of the finite scale can be a consequence of back-reaction, that is, the particle modifies the environment and drives the system away from criticality. However, the  model described here does not account for the effect of the particle on the environment. We leave for future research to test whether back-reaction implies transient subdiffusivity in this kind of systems. The latter scenario would not exclude the appearance of the infinite-time subdiffusion due to the emergence of a dynamical critical fixed point governing the particle-environment dynamics. 
 
\section{Ackowledgments}
 The authors acknowledge financial support from the ERC Advanced Grant OSYRIS, EU IP SIQS, EU PRO QUIC, EU STREP EQuaM (FP7/2007-2013, No. 323714), Fundaci\'o Cellex, the Spanish MINECO (SEVERO OCHOA GRANT SEV-2015-0522, FISICATEAMO FIS2016-79508-P), and the Generalitat de Catalunya (SGR 874 and CERCA/Program). MGP acknowledges European Commission (FP7-ICT-2011-7, Grant No. 288263) and the Spanish Ministry of Economy and Competitiveness ('Severo Ochoa' Programme for Centres of Excellence in R\&D SEV-2015-0522, and FIS2014-5617-R). CM acknowledges funding from the Spanish Ministry of Economy and Competitiveness (MINECO) and the European Social Fund (ESF) through the Ram\'on y Cajal program 2015 (RYC-2015-17896). GMG acknowledges funding from the Fundaci\'o Social LaCaixa.
 
 \appendix 
 \section{Numerical assessment of the number of steps in a domain}
 \label{sec:appA}

\begin{figure}
\includegraphics[width=0.95\columnwidth]{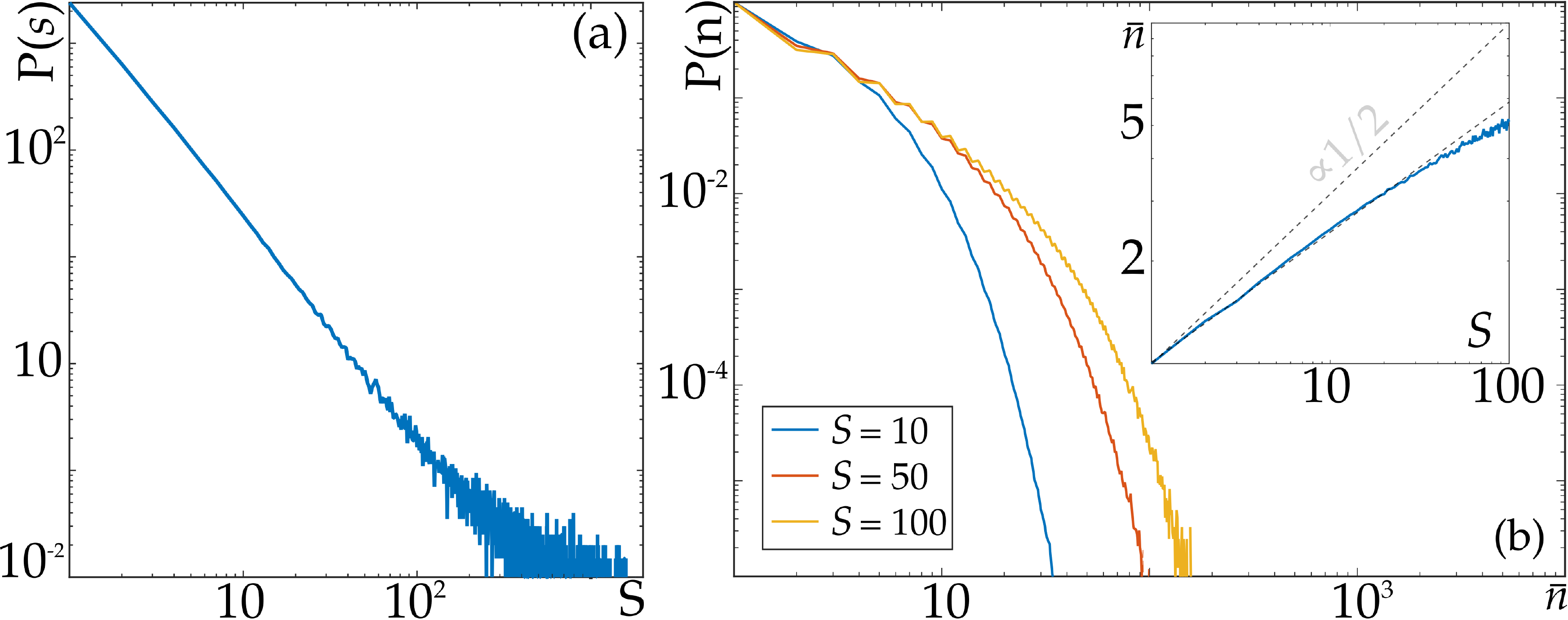}
\caption{  Distribution of the domain sizes and of the average number of steps $\overline{n}$ in a domain of $S$ obtained from Monte Carlo simulations.
 (a) Distribution of domain sizes for 200 patterns calculated at the critical temperature, $k_B T_c \approx 2.2691 J$, for $L=500$. (b) Numerically calculated probability distribution of the numbers of steps taken before  exiting a domain of size $S$ when the initial position  is in the border. To  do the calculations we select all the patterns of a given size $S$,  perform $X$ simulations of a particle starting at random positions in the border of the domain, and retain the time at which the particle  leaves the domain. Inset: average time $\overline n$  for leaving a domain of size $S$ in log-log scale, showing that   $\overline n$ grows approximately as $S^\mu$, with  $\mu\sim 0.3$.      \label{fig5} }
\end{figure}

In Fig.~\ref{fig5} (a) we illustrate the probability distribution of domain sizes, Eq.~(\ref{eq:distrS}). The distribution of sizes is calculated numerically  through Monte Carlo simulations, and the areas are extracted from 200 patterns calculated at  a temperature close to criticality ($k_B T_c \approx 2.2691 J$), with $L=500$. 
Using these patterns, we computed the average number of steps that a Brownian particle takes to leave a domain of size $S$, when starting from a random position on the border. To this end, we take all the domains of size $S$ and different shapes from our numerically calculated collection of patterns. We then compute the number of steps $n$ that a Brownian particle takes to leave the domain for many simulations and different initial conditions. We assume that $n$ follows a probability distribution, which we numerically calculate for many different $S$.   As shown in Fig.~\ref{fig5} (b), these distributions exponentially decay to zero for each $S$. We then calculate the average number of steps for leaving the domain, $\overline n$, from these distributions. We plot its logarithm as a function of the logarithm of $S$ in the inset of  Fig.~\ref{fig5} (b), where it presents a linear behavior with slope $\sim\!0.3$. It starts to deviate from linearity around $S=100$, which coincides with the domain size at which  fluctuations become important due to the lack of statistics [see Fig.~\ref{fig5} (a)]. Our numerical calculations with $L=1000$ showed similar results, but fluctuations started at a smaller $S$ because the number of patterns used was smaller. 
 
 \section{Numerical implementation}
 \label{sec:numimple}
 
To simulate the movement of a particle in an Ising environment we use a kinetic Monte Carlo algorithm. The waiting times are extracted from exponential distributions as in Eq.~(\ref{eq:waitingtimekappa}). The transition rates $w$ are equal for all the four neighbors of the site where the particle is at step $j$. Among the four possible displacements, we take the one corresponding to the shortest time~\cite{2014Toral} 
\begin{equation}
\label{eq:time}
t=\frac{\ln(U)}{w_{ij}}\overline\kappa,
\end{equation}
where $U$ is a random number between 0 and 1 and $\overline\kappa$ is obtained from the distribution of $\kappa$ (Eq.~(\ref{eq:distrK})). To obtain this value at each step, one needs to consider first the cumulative distribution function (CDF) associated to Eq.~(\ref{eq:distrSn}), which reads
\begin{equation}
\mathrm{CDF}_S=(\zeta-1)\int_1^{\overline S} S^{-\zeta}dS=1-\overline S^{1-\zeta}.
\label{eq:CDFS}
\end{equation}
From here, one obtains $\overline S=\left (1-\mathrm{CDF}_S\right )^{1/(1-\zeta)}$. In Eq.~(\ref{eq:CDFS}) one has to use transformation~(\ref{eq:transform}) to get
\begin{equation}
\mathrm{CDF}_\kappa=\eta(\zeta-1)\int_1^{\overline\kappa}\kappa^{\frac{1}{\eta(\zeta-1)}-1}d\kappa=1-\overline\kappa^{-\frac{1}{\eta}(\zeta-1)}.
\label{eq:CDFK}
\end{equation}
Therefore, $\overline\kappa=\left (1-\mathrm{CDF}_\kappa\right )^{\eta/(1-\zeta)}$, and subsequently 
\begin{equation}
\overline\kappa=\overline S^{\eta}.\label{eq:relkappaS}
\end{equation}
The latter relation proves that the sampled $\kappa$'s have the expected functional relationship with the sample size of the domain $S$ and it 
sets the value of $\kappa$ at step $j$ when the particle sits at a domain of size $S$, extracted from the corresponding probability distribution. 
At this point one can also find a relation between $\sigma$ in Eq.~(\ref{eq:distrK}), $\eta$, and $\zeta$. From Eq.~(\ref{eq:distrK}), one finds directly that the  CDF is
\begin{equation}
\mathrm{CDF}_\kappa=1-\overline\kappa^{1-\sigma},
\label{eq:CDFK1}
\end{equation}
and, by comparison with with Eq.~(\ref{eq:CDFK}) one obtains Eq.~(\ref{eq:beta}).

\end{document}